\documentclass[showpacs,preprint,aps]{revtex4}
\usepackage{graphicx}
\usepackage{dcolumn}
\usepackage{bm}
\begin{document}
\setcounter{page}{1}
\title 
{Breit Equation with Form Factors in the Hydrogen Atom}
\author
{F. Garc\'ia Daza, N. G. Kelkar and M. Nowakowski}
\affiliation{ 
Departamento de Fisica, Universidad de los Andes, 
Cra.1E No.18A-10, Bogota, Colombia}
\begin{abstract}
The Breit equation with two electromagnetic form-factors is studied to
obtain a potential with finite size corrections. This potential 
with proton structure effects includes apart from 
the standard Coulomb term, the Darwin term, retarded potentials, spin-spin 
and spin-orbit interactions corresponding to the fine and hyperfine structures 
in hydrogen atom. 
Analytical expressions for the hyperfine potential with form factors 
and the subsequent energy levels including the proton structure corrections 
are given using the dipole form of the form factors. Numerical results 
are presented for the finite size corrections in 
the 1S and 2S hyperfine splittings in the hydrogen atom, the Sternheim 
observable $D_{21}$ and the 2S and 2P hyperfine splittings in muonic hydrogen. 
Finally, 
a comparison with some other existing methods in literature is presented. 

\end{abstract}
\pacs{03.65.-w, 32.10.Fn, 13.40.Gp} 
\maketitle 
\section{Introduction}
The Breit equation \cite{Breit,Breit2} is a paradigm example of how one derives
coordinate potentials from Quantum Field Theory: an elastic scattering
amplitude, expanded in powers of $1/c^2$ and depending on the
three momentum transfer ${\bf q}$, gets Fourier transformed into
the coordinate space. The result is the potential $V(r)$. Famous examples
of this general principle include, among others, (i) 
the Casimir-Polder forces between neutral atoms 
as van der Waals forces at large distances from a two photon
exchange amplitude \cite{CasimirPolder}, (ii) 
the Uehling-Serber potential from one-loop vacuum polarization diagram 
\cite{Uehling}, (iii) the Feinberg-Sucher 
two neutrino exchange force \cite{FeinbergSucher}, (iv)
the microscopic potential of Nuclear Physics based on
$\sigma$, $\rho$ and $\omega$ exchanges  \cite{Nuclear} and (v)
quantum corrections to the Newtonian potential  \cite{Donoghue}. 
Other potentials are derived from pseudoscalar (Goldstone bosons) exchanges
\cite{pseudoscalar}, scalar Higgs exchanges (Higgsonium) \cite{grifols} 
and even
from Finite Temperature Quantum Field Theory which gives temperature
dependent potentials \cite{temperature}.
The span of the applications of these potentials, derived via the Fourier
transform of an elastic amplitude, ranges from atomic and molecular
physics up to nuclear-particle physics and even gravitation and cosmology 
\cite{cosmology}. 

The Breit equation follows the very same principle for elastic
$e^-\mu^+$, $e^+e^-$ (positronium), $e^-p$ (hydrogen) and
$\mu^-p$ (muonic hydrogen) amplitudes. The outcome is a potential
which goes beyond the Coulomb potential and includes the correct
expressions for fine structure, hyperfine structure and the Darwin
terms. To appreciate this fact let us mention that the fine and
hyperfine structure Hamiltonians in non-relativistic Quantum Mechanics
are derived by using semi-classical arguments \cite{griffiths} whereas
the Breit equation does so automatically in a systematic way without
reference to semi-classical arguments. This offers more insight into
semi-relativistic two body Hamiltonians. 
For instance, the standard Breit equation for two spin-$1/2$ particles 
can be extended to a spin-$1/2$ spin-$0$ system \cite{ourpaper1} necessary
for exotic atoms. Higher order terms in the $1/c^2$ expansion can be taken
into account and as we will show later, the finite size corrections
can be taken into account in a straightforward way. Indeed, the
finite size corrections to the potential have to do with the
way the photon interacts with an extended particle, i.e., with
a mild modification of the vertex for point-like particles (which is
to say that the modified vertex will now include the form-factors).
Hence, the modification of the vertices in the elastic amplitudes
leads after a Fourier transform to a potential which takes
into account the finite size corrections. This is a straightforward 
generalization of the standard Breit equation where the finite size
corrections are included in a natural way using a one-photon exchange diagram. 
The finite size corrections to the potential
can then be applied to calculate the finite size corrections to
the energy levels of the electronic or muonic hydrogen atom.
The necessity to do so is the ever increasing accuracy of the
theoretical QED corrections to the energy levels and the accurate
experimental results. 

The phenomenology of the hydrogen atom cannot be disentangled from the 
historical development of Quantum Mechanics (QM). Indeed, any progress in 
QM, going from non-relativistic to relativistic QM and finally to 
relativistic Quantum Field Theory, was accompanied by new corrections to the 
energy levels of the hydrogen atom and its cousin the muonic hydrogen
($\mu^- p$, 
where a negative muon replaces the electron in hydrogen). The hydrogen 
atom thus became a precision tool for testing predictions of Quantum 
Electrodynamics (QED) and more generally of electroweak theory, including 
symmetry considerations \cite{testqed}. The experimental techniques 
have reached an extraordinary precision \cite{accuexp}
which sometimes surpasses the theoretically 
calculated values expanded in the fine structure constant $\alpha$. 
For example, one of the most precisely measured quantities in physics 
is the hyperfine structure (hfs) of the ground state of hydrogen atom 
\cite{ramsey}, which is known to one part in $10^{13}$. 
In spite of this progress, there still remains one component, namely, the 
structure of the proton in the hydrogen atom which introduces uncertainties 
in the comparison of theories with experiment. Here, we 
demonstrate that the precision values calculated using QED 
get blurred by the nuclear finite size corrections (FSC). 
We focus in particular on these 
corrections in the hfs of hydrogen and $\mu^- p$. 
We use the theoretical framework of the Breit equation which in the case 
of the electromagnetic (EM) form factors taken at zero 
momentum transfer gives the standard hyperfine 
Hamiltonian. 
The frequencies of the hyperfine intervals for the ($1S$) and ($2S$) levels in 
hydrogen are first evaluated using the Breit equation with and without the 
EM form factors and compared with the precise experimental values. 
The proton structure correction to the 
Sternheim hyperfine interval \cite{sternheim} $D_{21}\, = 8 E_{hfs}(2S)
\, - \, E_{hfs}(1S)$ is evaluated and found to be much 
smaller than that obtained using the 
Zemach method \cite{zemach}. Finally, the FSC for 
the hfs in $\mu^- p$ which was used 
as an input for calculating the proton radius ($r_p$) 
in a recent accurate claim of the measurement of $r_p$ \cite{naturerp} are 
also evaluated. The above calculations show that the accurate measurements 
are not just limited due to the structure corrections and uncertainties in 
the proton form factors but also depend on the approach for the finite 
size corrections. 

In the next section, we introduce the framework of the Breit equation and 
derive the potential with finite size corrections due to the structure of 
the proton. After having presented the full Breit potential with form factors, 
in section III we focus on the part of the potential which corresponds to the 
hyperfine structure in hydrogen atom. The expressions for the corrections to 
the energy levels are evaluated using time independent perturbation theory. In 
section IV, we present 
the numerical results for hyperfine splitting in electronic 
($e^-p$) and muonic ($\mu^- p$) hydrogen and compare them with available 
data as well as results in literature obtained using other methods to 
incorporate the proton structure effects. The relevance of the results 
of the present work is summarized in the last section. 

\section{The Breit potential}
To start with, we briefly introduce the framework of the Breit equation. 
It involves an expansion of the elastic scattering amplitude, say, 
$e \,p \,\to\, e\, p$ in powers of $1/c^2$, thus having the 
advantage of systematically taking into account the relativistic corrections. 
Starting with the one photon exchange Feynman diagram for a system of two 
point-like spin 1/2 particles (like $e^-\mu^+$ for example), 
the full Hamiltonian for the $e^-p$ system considering point-like protons 
can be written as, 
\begin{equation}
\label{Hgeneral}
\hat{H}=\hat{H}_e^{(0)}+\hat{H}_p^{(0)}
+\hat{U}(\hat{\textbf{p}}_e,\hat{\textbf{p}}_p,\textbf{r})
\end{equation}
where it is convenient to split the potential into 
several parts (to be discussed below): 
\begin{equation}
\hat{U}(\hat{\textbf{p}}_e,\hat{\textbf{p}}_p,\textbf{r})=
\sum_{i=1}^{11}\hat{V}_i(\hat{\textbf{p}}_e,\hat{\textbf{p}}_p,\textbf{r}) . 
\end{equation}
The free Hamiltonian is expanded up to the same order in $1/c$ as 
the potential. In our case we have for example,  
\begin{equation}
\label{hfree}
 \hat{H}^0_{e,p} =
\frac{\hat{\textbf{p}}_{e,p}^2} {2 \, m_{e,p} }-\frac{\hat{\textbf{p}}_{e,p}^4}
{8 \, m_{e,p}^3c^2} . 
\end{equation}
The potential 
$\hat{U}(\textbf{p}_e,\textbf{p}_p,\textbf{q})$
in momentum space is obtained by writing the elastic amplitude $M_{fi}$ 
in terms of two-component spinors $w_{e,p}$, i.e., 
\begin{equation}
M_{fi}=-2 m_e\cdot 2 m_p(w_e^{'*}w_p^{'*})\hat{U}(\textbf{p}_e,\textbf{p}_p,
\textbf{q})(w_e w_p) . 
\end{equation} 
The Fourier transform of $\hat{U}(\textbf{p}_e,\textbf{p}_p,
\textbf{q})$ is the potential 
$\hat{U}(\hat{\textbf{p}}_e,\hat{\textbf{p}}_p,\textbf{r})$:
\begin{equation}
\hat{U}(\hat{\textbf{p}}_e,\hat{\textbf{p}}_p,\textbf{r})=
\int e^{i\textbf{q}.\textbf{r}}\:\hat{U}(\textbf{p}_e,\textbf{p}_p,
\textbf{q})\frac{d^3q}{(2 \,\pi)^3},
\end{equation}
where in the center of mass system we can identify 
$\hat{\textbf{p}}_e=-\hat{\textbf{p}}_p =- i \nabla$. 
The standard result for the potential in momentum space is \cite{LLbook}, 
\begin{eqnarray}
\label{potenqhyd}
 &&\hat{U}(\textbf{p}_e,\textbf{p}_p,\textbf{q})= 4\, \pi e^2
\Bigg(-\frac{1}{\textbf{q}^2}+\frac{1}{8m_e^2c^2}+\frac{1}{8m_p^2c^2}
+\frac{i\bm{\sigma}_p.(\textbf{q}\times\textbf{p}_p)}{4m_p^2c^2\textbf{q}^2}
-\frac{i\bm{\sigma}_e.(\textbf{q}\times\textbf{p}_e)}{4m_e^2c^2\textbf{q}^2}
\nonumber\\\nonumber\\& &
+\frac{\textbf{p}_e.\textbf{p}_p}{m_em_pc^2\textbf{q}^2}
-\frac{(\textbf{p}_e.\textbf{q})(\textbf{p}_p.\textbf{q})}{m_em_pc^2
\textbf{q}^4}
-\frac{i\bm{\sigma}_p.(\textbf{q}\times\textbf{p}_e)}
{2m_em_pc^2\textbf{q}^2}+\frac{i\bm{\sigma}_e.(\textbf{q}\times
\textbf{p}_p)}{2m_em_pc^2\textbf{q}^2}+\frac{\bm{\sigma}_e.
\bm{\sigma}_p}{4m_em_pc^2} \nonumber\\\nonumber\\& &-\frac{(\bm{\sigma}_e.
\textbf{q})(\bm{\sigma}_p.\textbf{q})}{4m_em_pc^2\textbf{q}^2}\Bigg),
\end{eqnarray}
and the individual terms of the potential in coordinate space 
come out to be, 
\begin{eqnarray}
\hat{V}_1&=&-\frac{e^2}{r}\label{v1a}\\
\hat{V}_2&=&\frac{\pi e^2}{2 m_e^2c^2} \delta(\textbf{r}), \label{v2a} \\
\hat{V}_3&=&\frac{\pi e^2}{2 m_p^2c^2} \delta(\textbf{r}),  \label{v3a}\\
\hat{V}_4&=&-\frac{e^2}{4 m_p^2c^2}\frac{(\textbf{r}
\times\hat{\textbf{p}}_p)}{r^3}.\bm{\sigma}_p, \label{v4a} \\
\hat{V}_5&=&\frac{e^2}{4 m_e^2c^2}\frac{(\textbf{r}\times
\hat{\textbf{p}}_e)}{r^3}.\bm{\sigma}_e,  \label{v5a}\\
\hat{V}_6&=&\frac{e^2}{m_e m_p c^2}\frac{\hat{\textbf{p}}_e.
\hat{\textbf{p}}_p}{r},\label{v6a}\\
\hat{V}_7&=&-\frac{e^2}{2 m_e m_p c^2r}\left( \hat{\textbf{p}}_e.
\hat{\textbf{p}}_p-\frac{\textbf{r}.(\textbf{r}.\hat{\textbf{p}}_e)
\hat{\textbf{p}}_p}{r^2}\right),\label{v7a}\\
\hat{V}_8&=&\frac{e^2}{2 m_e m_p c^2}\frac{(  
\textbf{r}\times\hat{\textbf{p}}_e)}{r^3}.\bm{\sigma}_p, \label{v8a}\\
\hat{V}_9&=& -\frac{e^2}{2 m_e m_p c^2}\frac{(  \textbf{r}\times
\hat{\textbf{p}}_p)}{r^3}.\bm{\sigma}_e, \label{v9a}\\
\hat{V}_{10}&=& \frac{\pi e^2}{m_e m_p c^2}\left(\bm{\sigma}_e.
\bm{\sigma}_p\right) \delta(\textbf{r}),\label{v10a}\\
\hat{V}_{11}&=&-\frac{e^2}{4 m_e m_pc^2}
\left( \frac{\bm{\sigma}_e.\bm{\sigma}_p}{r^3}-3
\frac{(\bm{\sigma}_e.\textbf{r})(\bm{\sigma}_p.
\textbf{r})}{r^3}+\frac{4 \pi}{3}\bm{\sigma}_e.\bm{\sigma}_p 
\delta(\textbf{r})\right).\label{v11a} 
\end{eqnarray}
Here $e^2= \alpha$. 
The meaning of the terms is as follows:
\begin{itemize}
\item
$\hat{V}_1$ is obviously the Coulomb potential
\item 
$\hat{V}_2$ and $\hat{V}_3$ are the Darwin terms which are also present in the
Dirac equation \cite{Darwin}
\item
$\hat{V}_6$, $\hat{V}_7$ are called retarded potentials
\item
$\hat{V}_5$ and 
$\hat{V}_9$ are the spin-orbit interaction terms which give rise to the fine
structure; $\hat{V}_4$ and $\hat{V}_8$ are also a spin-orbit
terms, however, traditionally
added to the hyperfine part \cite{bethe}
\item
$\hat{V}_{10}$, $\hat{V}_{11}$
are the standard terms of the hyperfine Hamiltonian.
\end{itemize}
It is clear that the hyperfine part given above 
is valid only for point-like particles.
For the electron-proton system we miss the correct magnetic moment 
of the proton.
Indeed, for the hydrogen atom, 
the standard hyperfine potential 
as in text books \cite{griffiths} reads in natural units, $\hbar=c=1$, 
\begin{equation}\label{hfsbook}
\hat{V}_{\rm hfs}= 
{\alpha \over m_e m_p c^2} {g_p \over 2}\left[\frac{3({\bf S}_e\cdot 
\hat{\bf r})({\bf S}_e\cdot \hat{\bf r})-{\bf S}_e\cdot {\bf S}_p}{r^3} \, 
+\, {8 \pi \over 3} {\bf S}_e \cdot {\bf S}_p 
\, \delta^3({\bf r}) \right]
\end{equation}
with $g_p=5.58$. 
Replacing \[{\bf S}_{e,p} = \mbox{\boldmath$\sigma$}_{e,p}/2\] and 
comparing with Eq. (\ref{v11a}), one can see that $\hat{V}_{11}$ in 
(\ref{v11a}) gives the hyperfine potential, however, without the 
factor $g_p/2$. The missing factor in the 
Breit Hamiltonian of Eq. (\ref{v11a}) is part of the second form-factor of
the proton, namely,   
\begin{equation}
[1+\kappa_p]=[1+F_2^p(0)]=g_p/2 .  
\end{equation}
The purpose of this small exercise is to bring
to the reader's attention that
for point-like particles, $F_2(0)$ is very small (and hence 
$g_p \approx 1$ in the Breit potential in (\ref{v11a})), as this quantity
arises from one and more loops (anomalous magnetic moment).
The hyperfine potential in (\ref{hfsbook}) already includes part of the 
finite size corrections (in the form of $g_p = 2 [1+F_2^p(0)]$ ).
To remedy the situation for the $e^-p$ system in the Breit equation
we enlarge it by allowing the electromagnetic form-factors in
the vertex for the proton (and for the sake of generality also
for the electron) as follows: 
\begin{equation}
\bar{u}(p')\:\Gamma^\mu(p',\:p)\: u(p)=\bar{u}(p')\:
\Bigg(\gamma^\mu F_1(q^2)+
\frac{i}{2 m_p} F_2(q^2)\sigma^{\mu\nu}q_\nu\Bigg)\: u(p) . 
\end{equation}
Doing so, we obtain the following potential in momentum space \cite{ourpaper1}
\begin{eqnarray}
&&\hat{U}(\textbf{p}_e,\textbf{p}_p,\textbf{q})=4\pi e^2\Bigg[F_1^eF_1^p\Bigg
(-\frac{1}{\textbf{q}^2}+\frac{1}{8m_e^2c^2}
+\frac{1}{8m_p^2c^2}+\frac{i\bm{\sigma}_p.(\textbf{q}\times\textbf{p}_p)}
{4m_p^2c^2\textbf{q}^2}
-\frac{i\bm{\sigma}_e.(\textbf{q}\times\textbf{p}_e)}{4m_e^2c^2\textbf{q}^2}	\nonumber\\\nonumber\\	& &
 +\frac{\textbf{p}_e.\textbf{p}_p}{m_em_pc^2\textbf{q}^2}-\frac
{(\textbf{p}_e.\textbf{q})(\textbf{p}_p.\textbf{q})}{m_em_pc^2\textbf{q}^4}
-\frac{i\bm{\sigma}_p.(\textbf{q}\times\textbf{p}_e)}{2m_em_pc^2\textbf{q}^2}
+\frac{i\bm{\sigma}_e.(\textbf{q}\times\textbf{p}_p)}
{2m_em_pc^2\textbf{q}^2}+\frac{\bm{\sigma}_e.\bm{\sigma}_p}{4m_em_pc^2}	
\nonumber\\\nonumber\\& &
 -\frac{(\bm{\sigma}_e.\textbf{q})(\bm{\sigma}_p.\textbf{q})}
{4m_em_pc^2\textbf{q}^2}\Bigg)+F_1^eF_2^p\Bigg(\frac{1}{4m_p^2c^2}
+ \frac{i\bm{\sigma}_p.(\textbf{q}\times\textbf{p}_p)}{2m_p^2c^2\textbf{q}^2}  
- \frac{i\bm{\sigma}_p.(\textbf{q}\times\textbf{p}_e)}{2m_em_pc^2\textbf{q}^2}
-\frac{(\bm{\sigma}_e.\textbf{q})(\bm{\sigma}_p.\textbf{q})}{4m_em_pc^2
\textbf{q}^2} 	
\nonumber\\\nonumber\\& &+ \frac{\bm{\sigma}_e.\bm{\sigma}_p}{4m_em_pc^2}\Bigg)
 +F_2^eF_1^p\Bigg(\frac{1}{4m_e^2c^2} - \frac{i\bm{\sigma}_e.
(\textbf{q}\times\textbf{p}_e)}{2m_e^2c^2\textbf{q}^2}  
+ \frac{i\bm{\sigma}_e.(\textbf{q}\times\textbf{p}_p)}{2m_em_pc^2\textbf{q}^2}
-\frac{(\bm{\sigma}_e.
\textbf{q})(\bm{\sigma}_p.\textbf{q})}{4m_em_pc^2\textbf{q}^2} \nonumber\\\nonumber\\& &	
+ \frac{\bm{\sigma}_e.\bm{\sigma}_p}{4m_em_pc^2}\Bigg)+	
F_2^eF_2^p\Bigg(\frac{\bm{\sigma}_e.\bm{\sigma}_p}{4m_em_pc^2}
-\frac{(\bm{\sigma}_e.\textbf{q})(\bm{\sigma}_p.\textbf{q})}{4m_em_pc^2
\textbf{q}^2}\Bigg)
 \Bigg] .\label{potFF}
\end{eqnarray}
If one now takes all form-factors at zero momentum transfer: $F_1^e(0)=1$, 
$F_2^e(0)=\kappa_e=1159.6521811(7)\times 10^{-6}$,
$F_1^p(0)=1$, 
$F_2^p(0)=\kappa_p=1.792847351(2)$, the potentials are:  
\begin{equation}
\hat{V}_{\rm fine}=\frac{e^2}{4m_e^2c^2}\Bigg(\frac{(\textbf{r}\times\hat{\textbf{p}}_e)}{r^3}(1+2\kappa_e)
-\frac{2m_e}{m_p}\frac{(  \textbf{r}\times\hat{\textbf{p}}_p)}{r^3}(1+\kappa_e)\Bigg).\bm{\sigma}_e,\label{fineLE}
\end{equation}
for the fine structure and 
\begin{eqnarray}
 &&\hat{V}_{\rm hfs}=\frac{e^2}{2m_em_pc^2}\Bigg(\frac{(  \textbf{r}\times\hat{\textbf{p}}_e)}{r^3}(1+\kappa_p)
-\frac{m_e}{2m_p}\frac{(\textbf{r}\times\hat{\textbf{p}}_p)}{r^3}(1+2\kappa_p)\Bigg).\bm{\sigma}_p\nonumber\\& &%
 +(1+\kappa_e)(1+\kappa_p)\frac{e^2}{4m_em_pc^2}\Bigg( -
\frac{\bm{\sigma}_e.\bm{\sigma}_p}{r^3}+3\frac{(\bm{\sigma}_e.
\hat{\textbf{r}})(\bm{\sigma}_p.\hat{\textbf{r}})}{r^3}+\frac{8\pi}{3}\bm{\sigma}_e.
\bm{\sigma}_p\delta(\textbf{r})\Bigg).\label{hyperfineEL}
\end{eqnarray}
for the hyperfine part where we have included some part of the spin-orbit interaction terms.
With $e^2=\alpha$ and 
neglecting $\kappa_e$ and the term suppressed by $m_e m_p$,  
the above fine structure term due to spin orbit coupling (Eq. \ref{fineLE}) 
reduces to 
\begin{equation}
\hat{V}^{book}_{\rm fine}=\frac{\alpha}{4m_e^2c^2}
\Bigg(\frac{(\textbf{r}\times\hat{\textbf{p}}_e)}{r^3}\Bigg).
\bm{\sigma}_e \, =\, \frac{\alpha}{2m_e^2c^2}
\Bigg(\frac{\bf{L} \cdot \bf{S}_e}{r^3}\Bigg)
\end{equation}
found in books \cite{griffiths} (note that in \cite{griffiths} the notation 
is slightly different and $e^2 = 4 \pi \epsilon_0 \alpha$) . 
The hyperfine
potential after neglecting the $\kappa_e$ term, the spin orbit term and 
the one 
suppressed by $m_p^2$, reduces to Eq.(\ref{hfsbook}) 
which is also a standard result found in books \cite{griffiths}. If we 
retain the spin orbit term (not suppressed by $m_p^2$) however, we recover 
\begin{equation}
\hat{V}^{book}_{\rm hfs}=\frac{\alpha}{m_em_pc^2}\, {g_p \over 2}
\left [  \frac{(  \textbf{r}\times\hat{\textbf{p}}_e)}{r^3} \cdot 
\bf{S}_p \, 
+ \frac{3({\bf S}_e\cdot 
\hat{\bf r})({\bf S}_e\cdot \hat{\bf r})-{\bf S}_e\cdot {\bf S}_p}{r^3} \, 
+\, {8 \pi \over 3} {\bf S}_e \cdot {\bf S}_p 
\, \delta^3({\bf r}) \right]
\end{equation}
which in the absence of any proton structure, i.e., dropping the 
factor $g_p/2$ is another text book result 
as in \cite{bethe}.

We emphasize again that we have obtained this agreement after retaining 
the proton form factors at $q^2=0$, i.e., $F_1^p(0)=1$ and 
$F_2^p(0) = \kappa_p$ which is not a small number. 
$F_2^p(0)$  would be tiny if the proton would be a point-like particle. 
This implies that the hyperfine potential in books, necessarily 
includes parts of the finite size corrections (in the form 
of $F_2^p(0)$). It is now straightforward to
generalize the result to take into account the full finite size 
corrections to the
potentials by keeping the full $q^2$ dependence, i.e., using 
$F_1^p(q^2)$ and $F_2^p(q^2)$. 

\section{Hyperfine splitting in $e^-p$ and $\mu^-p$ atoms} 
Having discussed the complete Breit potential with electromagnetic form 
factors, we shall now focus on the part of the potential which gives rise 
to the hyperfine splitting in hydrogen atom. We do this with the objective 
of evaluating the finite size corrections (FSC) to the hyperfine energy 
levels and comparing them with precision data which is available for the 
$l=0$ and $l=1$ levels.
As outlined in the previous section, normally terms suppressed by small 
factors arising due to large hadron masses in the denominator are neglected. 
In the calculations to follow, we shall neglect the term proportional to 
$1/m_p^2$ in the hyperfine potential. The factor $1/m_p^2$ appears in the 
spin-orbit term which in any case is either small (for $l=1$) or not 
relevant (for $l=0$). 
\subsection{Hyperfine potential with form factors} 
The hyperfine 
potential in momentum space contains the proton spin-orbit interaction and the 
proton-electron (or proton-muon 
in the case of $\mu^- p$) spin-spin interaction which 
are responsible for the hyperfine 
structure. 
\begin{eqnarray}\label{potqhfs}
\hat{U}({\bf q})_{hfs}  = \pi \alpha \biggl [ \, 
{(\mbox{\boldmath$\sigma$}_X \cdot 
\mbox{\boldmath $\sigma$}_p) \over m_X\, m_p c^2}  - 
{(\mbox{\boldmath$\sigma$}_X \cdot {\bf q}) \cdot (
\mbox{\boldmath$\sigma$}_p \cdot {\bf q}) \over 
m_X\, m_p c^2  {\bf q}^2} \, \biggr ] 
[ (F_1^X + F_2^X) (F_1^p({\bf q}^2)\, + F_2^p({\bf q}^2))] 
\\ \nonumber 
- \biggl [ \,(2 \pi \alpha)\,  
{i \mbox{\boldmath$\sigma$}_p ({\bf q} \times {\bf p}_X) \over m_X\, m_p c^2 
{\bf q}^2}\, 
\biggr ] \, [F_1^X\, (F_1^p({\bf q}^2)\,+ \, F_2^p({\bf q}^2))] \, +\, 
\biggl [ \,(2 \pi \alpha)\,{i \mbox{\boldmath$\sigma$}_p ({\bf q} \times 
{\bf p}_p) \over m_p^2 c^2 
{\bf q}^2}\, \biggr ] \, F_1^X F_2^p({\bf q}^2)) \, ,
\end{eqnarray} 
with, $X = e, \mu$ for the electron or muon, $F_1^X\, =\, 1$ and 
$F_2^X\, =\, \kappa_X$ (the anomalous magnetic moment). $F_1^p({\bf q}^2)$ and 
$F_2^p({\bf q}^2)$ are the two electromagnetic 
form factors of the proton. Early experiments on electron-proton elastic 
scattering showed that the cross sections can be written in terms of 
two form factors, $G_E^p(q^2)$ and $G_M^p(q^2)$, where $q^2 = q_0^2 - 
{\bf q}^2$. These  
Sachs form factors \cite{punjabi} can be interpreted 
in the Breit frame to be the Fourier transforms of 
the charge ($\rho_C$) and magnetization ($\rho_M$) 
distributions in the proton: 
\begin{eqnarray}
G_E^p(q^2) &=& \int \, \rho_{C}({\bf r})\, e^{-i{\bf q} \cdot {\bf r}} 
\, d{\bf r}\\ \nonumber
G_M^p(q^2) &=&\mu_p 
\int \, \rho_{M}({\bf r})\, e^{-i{\bf q} \cdot {\bf r}} 
\, d{\bf r} . 
\end{eqnarray}
They can be approximated fairly well by a dipole form 
\cite{bosted} as follows:
\begin{equation}
G_D(q^2) = {1 \over (1 + q^2/m^2)^2} \approx G_E^p(q^2) \approx 
{G_M^p(q^2)\over \mu_p} , 
\end{equation}
where $m^2 = 0.71$ GeV$^2$. 
They are related to $F_1^p(q^2)$ and $F_2^p(q^2)$ as, 
\begin{eqnarray}
G_E^p(q^2) = F_1^p(q^2) + {q^2\over 4 m_p^2c^2} F_2^p(q^2) ,\\ \nonumber
G_M^p(q^2) = F_1^p(q^2) +  F_2^p(q^2) . 
\end{eqnarray}
Picking up the first terms in the round brackets following $F_1^e F_1^p$ and 
$F_1^e F_2^p$ in the Breit potential in 
(\ref{potFF}) and putting them together, one gets (using 
$q^2 \approx - {\bf q}^2$ which is the standard non-relativistic approximation 
to derive potentials from amplitudes \cite{LLbook}),  
$$- {4 \pi \alpha \over {\bf q}^2} \biggl [ F_1^p({\bf q}^2) - 
{\bf q}^2 
{F_2^p({\bf q}^2) \over 4 m_p^2 c^2} \biggr ] = - {4 \pi \alpha \over {\bf q}^2} 
G_E^p({\bf q}^2),$$
which is the standard definition in ${\bf q}$-space for the Coulomb 
potential due to the finite size of the proton.
Thus, 
\begin{eqnarray}\label{f1f2}
F_1^p({\bf q}^2)\, =\, {1 \over (1\, +\, {\bf q}^2/m^2)^2} \, 
\biggl [ \, 1 \, +\, 
\kappa_p\, {{\bf q}^2 \over 4\, m_p^2\, +\, {\bf q}^2} \, \biggr ], \\ 
\nonumber
F_2^p({\bf q}^2)\, =\, {1 \over (1\, +\, {\bf q}^2/m^2)^2} \, {4\, m_p^2\, 
\kappa_p \over 
4\, m_p^2\, +\, {\bf q}^2} \, ,
\end{eqnarray}
where $m_p$ is the mass of the proton, ($1\, + \,\kappa_p$)  
= $\mu_p$ = 2.793 its 
magnetic moment and $m^2$ = 0.71 GeV$^2$. 

The nuclear finite size corrections (FSC) mentioned in the beginning are 
thus rooted 
in the two form factors $F_1^p$ and $F_2^p$. The nuclear uncertainty 
in general can be traced back to (i) different methods used to calculate the 
FSC (the Breit equation formalism being one such method which,
however, is derived directly from quantum field theory) and (ii) the 
limited precision in the measurement of the two proton form factors. 
In the present work, we shall restrict ourselves to the dipole form of 
the form factors given above. The reason for using this approximation is that 
(i) the potentials as well as corrections to the energy levels can be 
evaluated analytically and (ii) the objective of this work is to present the 
new method of the Breit equation for calculating the FSC and not 
explore the uncertainties introduced due to the use of different 
parametrizations \cite{alberico} of the form factors.

Replacing (\ref{f1f2}) in (\ref{potqhfs}) and taking the Fourier transform, 
the potential for the hyperfine structure in $r$-space is given as, 
\begin{equation}\label{potrhfs}
\hat{V}_{hfs}(r) = {\alpha \mu_p\over 4 r^3 m_X m_p c^2}  
\biggl [ \mu_X \biggl \{ {3 (\mbox{\boldmath$\sigma$}_X 
\cdot \hat{\bf r} ) (\mbox{\boldmath$\sigma$}_p \cdot \hat{\bf r})} 
f_1(r)\, -\, {\mbox{\boldmath$\sigma$}_X \cdot \mbox{\boldmath$\sigma$}_p} 
f_2(r)\biggr \} 
+ 2 {{\bf L} \cdot \mbox{\boldmath$\sigma$}_p} f_3(r) \biggr ],  
\end{equation}
where, $\mu_X \, =\, 1\, +\, \kappa_X$,
$$f_1(r)\, =\, 1\, -\, e^{-mr} (1\,+\, mr)\, -\, {m^2 r^2 \over 6} 
\, e^{-mr} \, (3\, +\, mr),$$ 
$$f_2(r) = f_1(r) - (m^3 r^3/3)e^{-mr}\,\, {\rm and},$$ 
$$f_3(r) = 1\, -\, e^{-mr} (1\,+\, mr)\, 
-\, {m^2 r^2 \over 2} \, e^{-mr}.$$
For $l=0$ and point like 
protons, i.e., replacing $F_1^p\, =\, 1$ and
$F_2^p\, =\, \kappa_p$,  
the Fourier transform of the potential (\ref{potqhfs}) leads to the 
standard text book potential \cite{LLbook} discussed in the previous 
section. Thus 
\begin{equation}\label{potrhfspoint}
V^{point}_{hfs}(r)\, =\, {\alpha \over 4 m_X m_p c^2} \, 
\mu_X\, \mu_p\, \biggl [\, {3 (\mbox{\boldmath$\sigma$}_X 
\cdot \hat{\bf r} ) \,(\mbox{\boldmath$\sigma$}_p \cdot \hat{\bf r})\over r^3} 
\, -\, {\mbox{\boldmath$\sigma$}_X \cdot \mbox{\boldmath$\sigma$}_p 
\over r^3} \, +\, {8\pi\over 3}\, 
\mbox{\boldmath$\sigma$}_X \cdot \mbox{\boldmath$\sigma$}_p\, \delta({\bf r})\, \biggr ]\, , 
\end{equation}
is similar to (\ref{hyperfineEL}) except for the first two terms 
in (\ref{hyperfineEL}) 
corresponding to the spin-orbit coupling which is absent in the $l=0$ case. 
Since in the ground state ($l=0$), the wave function is spherically symmetric, 
the expectation value of the first two terms in (\ref{potrhfspoint}) vanishes 
and the third term with the delta function contributes to 
the energy $E^{point}_{hfs}$ of the $l=0$ hyperfine levels. 
However, once we include the form factors of the proton which are 
smooth functions of $q^2$, there is no singularity or a delta function term  
and both terms in the curly bracket of (\ref{potrhfs}) 
contribute to $E_{hfs}$ for $l = 0$. 
Eq. (\ref{potrhfs}) contains the standard hyperfine structure 
plus terms involving the FSC and anomalous magnetic moments of the 
electron or muon. 
\subsection{Finite size corrections to the energy levels}
For any $n$ and 
$l = 0$ for example, in case of the hydrogen atom, 
we need to evaluate the expectation 
value, 
$\langle\, [3\, \mbox{\boldmath$\sigma$}_e \cdot \hat{\bf r}) 
(\mbox{\boldmath$\sigma$}_p \cdot 
\hat{\bf r}) 
\, f_1(r) \, - \, (\mbox{\boldmath$\sigma$}_e \cdot \mbox{\boldmath$\sigma$}_p)
\, f_2(r)] / r^3\, 
\rangle $. Using the fact that $\int \, ({\bf a} \cdot \hat{\bf r}) \, 
({\bf b} \cdot \hat{\bf r}) \, {\rm sin}\theta\, d\theta \,d\phi\, =\, 
(4\pi /3) \, ({\bf a} \cdot {\bf b})$, the energy of the hyperfine levels 
as evaluated from (\ref{potrhfs}) for any $n$ and $l = 0$ is given as, 
\begin{eqnarray}\label{energyhfsn}
E^{n,l=0}_{hfs}\, =\, {\alpha \over 4 m_e m_p c^2} \, (1\, +\, \kappa_e) \, 
\mu_p \, \langle \mbox{\boldmath$\sigma$}_e \cdot \mbox{\boldmath$\sigma$}_p 
\rangle \, 
{m^3 \over 3} \, \biggl ({2 \over na} \biggr )^3 \, {\Gamma(n+1) \over 
2^{2n-1}\, n (n!)} \\ \nonumber 
\times \, \displaystyle{\sum_{k=0}^{n-1}}\,\left(
\begin{array}{c}
2 n \, -\,2 k\, -\, 2
\\
n \, -\, k\, -\, 1
\end{array}
\right )\, 
{\Gamma(2k+3) \over k! \, \Gamma(k+2)} \, \biggl ( \, {na \over 2\, +\, mna} 
\, \biggr )^3\, _2F_1\biggl (\, -2k,\,3; \, 3;\, {4 \over 2 + mna}\biggr )\, , 
\end{eqnarray}
where, 
$a=1/(\alpha m_r)$ is the Bohr radius with the reduced mass $m_r$ and 
$_2F_1$ is the confluent hypergeometric function of the second kind. 
Replacing the series $$_2F_1 (a,b;c;z) \,=\,1\, +\, (ab/c)(z/1!)\, +\, 
[a (a+1) b (b+1)]/[c(c+1)](z^2/2!)\, +\, ...... $$ in 
(\ref{energyhfsn}) and evaluating for $n = 1$ and $n = 2$, 
the energies of the $(1S)$ and $(2S)$ hyperfine levels in 
hydrogen are given as, 
\begin{eqnarray}\label{energyhfs12}
E^{1S}_{hfs}& =&\alpha^4\, \biggl (\, {m_r^3\over 4 m_e m_p c^2} \, \biggr )\, 
\mu_e\, \mu_p\, \langle \mbox{\boldmath$\sigma$}_e \cdot 
\mbox{\boldmath$\sigma$}_p 
\rangle \,{8 \over 3}\, \biggl [ \, 1\, -\, 3 \biggl ({2\over ma} 
\biggr ) \, +\, 6 \biggl ( {2 \over ma} \biggr )^2\,+\,...\biggr ] 
\\ \nonumber
E^{2S}_{hfs}& =&\alpha^4\, \biggl (\, {m_r^3\over m_e m_p c^2} \, \biggr ) \, 
\mu_e\, \mu_p\, \langle \mbox{\boldmath$\sigma$}_e \cdot 
\mbox{\boldmath$\sigma$}_p 
\rangle \,{1 \over 3}\, \biggl [ \, 1\, -\, 6 \biggl ({1\over ma} 
\biggr ) \, +\, 21 \biggl ( {1 \over ma} \biggr )^2\, +\,...\biggr ] 
\, . 
\end{eqnarray}
These are the energies of the hyperfine levels as obtained from the Breit 
equation for the $e\, p\,\to\, e\, p$ amplitude including the proton 
form factors $F_1^p({\bf q}^2)$, $F_2^p({\bf q}^2)$ (
as in Eq. (\ref{f1f2})) and 
$F_1^e\,=\, 1$, $F_2^e\, =\, \kappa_e$ at the two vertices. In the absence 
of the proton FSC, i.e., for $m \to \infty$ in 
(\ref{energyhfs12}) and not taking into account the correction due to the 
anomalous electron magnetic moment (i.e. $\kappa_e \to 0$), 
the energies in Eq. (\ref{energyhfs12}) 
go over to the expressions in standard text books 
\cite{griffiths,bethe}. In the next section we shall present a comparison 
of the corrections evaluated using the above equations with accurate data 
available for the $1S$ and $2S$ hyperfine levels. 

In \cite{naturerp}, the size of the proton was deduced 
from a measurement of the muonic Lamb shift. The total predicted 
2S$^{f=1}_{1/2}$ - 2P$^{f=2}_{1/2}$ energy difference is a sum of 
contributions from the fine and hyperfine splittings among others. The authors 
claimed a high accuracy on the radius deduced from this energy difference 
and concluded that the Rydberg constant has to be shifted by -110 kHz/c. 
In view of these results it becomes important to estimate the proton 
FSC to the hyperfine splitting relevant to this transition 
in the most sophisticated way possible, within a 
given approach. Furthermore, one should take into account the 
uncertainties arising due to the use of different approaches. 
Hence, we also evaluate corrections 
for the 2S and 2P hyperfine levels in muonic hydrogen using the Breit 
potential with form factors. 
The energy $E^{2P}_{hfs}$ evaluated using the potential in 
Eq. (\ref{potrhfs}) for the triplet spin states is, 
\begin{eqnarray}\label{energyhfs2p}
E^{2P}_{hfs} &=& {\alpha \mu_p \over 24 a ^3 m_{\mu} m_p c^2} 
\biggl [  
{-\mu_{\mu} \over 4 j (j+1)} \biggl (j(j+1)-l(l+1)-{3\over 4} \biggr)
\biggl ( f(f+1)-j(j+1)-{3\over 4} \biggr ) \\ \nonumber
&&\times [1 - g_1(ma) - 2 g_2(ma)] + g_2(ma)
+ {1 \over 4 j (j+1)} \biggl (j(j+1)+l(l+1)-{3\over 4}\biggr ) \\ \nonumber
&&\times \biggl ( f(f+1)-j(j+1)-{3\over 4}\biggr ) [1 - g_1(ma)] \biggr ] 
\end{eqnarray}
where,
$$
g_1(ma) =  \tilde{f}^2(ma) + 2 m a 
\tilde{f}^2(ma) + 3 m^2 a^2 
\tilde{f}^4(ma)$$ and
$$g_2(ma) = 2 m^3 a^3 \tilde{f}^5(ma) \, \, 
{\rm 
with} \, \, \tilde{f}(ma)=1/(1+ma).$$
The quantum numbers $f$, $j$ and $l$ refer to the total angular momentum 
of the system, $f = j + s_p$, the electron total angular momentum, 
$j = l + s_e$ and the orbital angular momentum $l$. 
The expressions for any $n$ and $l$ contain the hypergeometric functions 
$_2F_1$ as before. Since these expressions are extremely lengthy, we present 
them in the appendix and here we write the result in the form of 
Eq.(\ref{energyhfs2p}) using a truncated expansion of
$_2F_1$ as done before Eqs (\ref{energyhfs12}). 

\section{Results and discussions}
The frequencies of the $l=0$ hyperfine intervals in hydrogen atom have been 
measured quite accurately. The precise value of 
the ($2S$) splitting in literature is 
177556834.3(6.7) kHz \cite{kolachev} and that of the ($1S$) 
is 1420405.751768(1) kHz \cite{karshenrep}. 
In Table I we present the 
corrections to the frequencies 
calculated using Eqs (\ref{energyhfs12}) for the hyperfine 
energies. 
Denoting the hyperfine energy with $\kappa_e \to 0$ 
and $m \to \infty$ as $E^0$, one with only $m \to \infty$ as $E^{no\,FSC}$, 
the FSC correction introduced due to the proton form factors is denoted as 
$\Delta E^{FSC}\, =\, E^{nS}_{hfs} \, -\, E^{no\,FSC}$. The correction 
due to the proton form factors as well as the anomalous magnetic moment of the 
electron is denoted as $\Delta E^{corr}\, =\, E^{nS}_{hfs} \, -\, E^0$. 

\begin{table}[ht]
\caption{ Frequencies in kHz of hyperfine splittings in hydrogen atom.} 
\label{tab:1}
\begin{tabular}{|c|c|c|c|}\hline
(nl) &$f^0$  &$\Delta f^{FSC}$& $\Delta f^{corr}$ \\
     &       & due to proton structure& due to $\kappa_e$ and proton structure\\
\hline
(1S) &1418840.09  &-37.696 &1607.665  \\ \hline
(2S) &177355.01 &-4.712  & 200.958 \\ \hline
\end{tabular} 
\end{table}
One can see that the FSC 
are large and relevant considering the accuracy of the experimental numbers. 
The FSC are often included via a multiplicative factor containing form factors 
\cite{zemach,pach,friar} 
or more simply by introducing a correction depending on the
proton radius \cite{itzyk}. 
One expects \cite{karshenPLB} 
the difference $D_{21} \, =\, 8\, E^{2S}_{hfs} \, -\,  E^{1S}_{hfs}$ to 
be free of proton finite size corrections. From Eqs (\ref{energyhfs12}) one 
can see that this difference is finite but small and we find 
$\Delta D_{21} \, =\, -0.0833$ Hz. 

\subsection{Comparison with other methods}
The need to include structure corrections to the hyperfine energy levels 
due to the finite size of 
the proton in the hydrogen atom was noticed about 50 years ago. After the 
pioneering work of Zemach \cite{zemach} many other calculations followed. 
In what follows, 
we first summarize the main aspects of the Breit potential method and
then present a comparison with other methods in literature.
\subsubsection{Breit potential method}
This method follows the standard method
of quantum field theory to calculate a potential as 
the Fourier transform of an elastic amplitude. 
As mentioned in the introduction, it is a well documented method and
widely used in different branches of physics. 
The finite size corrections (FSC) 
to the potential stem from replacing the point-like
vertex in the one photon exchange amplitude 
(the latter without form factors gives rise to the 
hyperfine Hamiltonian in the standard Breit equation)
by the standard vertex which includes form factors. The result
thus generalizes the standard Breit equation to encode all necessary
FSC to the potential. It should be emphasized again that the
standard Breit equation gives all operators (fine and hyperfine)
which appear in the Hamiltonian for the hydrogenic atom. It is therefore
true to say that the fine and hyperfine operators come directly from
quantum field theory. A perturbation of these operators 
by the electromagnetic form factors seems then to be the most natural way
to include FSC. In the second step we use
the unperturbed Coulomb wave functions in the time-independent
perturbation theory to arrive at the FSC to the energy levels. Thus the 
correction to the hyperfine energy level, $\Delta E_{hfs}$, is given by, 
$\Delta E_{hfs} = E_{hfs} - E^{noFSC}_{hfs}$, where  
\begin{equation}\label{encorr}
E_{hfs} = \int \, \Psi^*({\bf r}) \, V_{hfs}(r) \, \Psi({\bf r}) \, 
d{\bf r}\, . 
\end{equation}
$\Psi({\bf r})$ is the usual hydrogen atom wave function 
(solution of the unperturbed Hamiltonian) and $V_{hfs}(r)$ is the 
hyperfine potential with form factors (as given in Eq.(\ref{potrhfs})). 
$E^{noFSC}_{hfs}$ is evaluated as in (\ref{encorr}), however, with the 
potential $V_{hfs}(r)$ replaced by $V^{point}_{hfs}(r)$. 
Note that this method involves a systematic expansion in $\alpha$.
This is to say, if an operator being part of the Hamiltonian without
FSC is of order $\alpha^n$, the corresponding FSC 
to the potential will also come out to be of the same order. 
\subsubsection{Zemach's original work} 
The approach of Zemach as presented in \cite{zemach} 
is very different from that of the Breit potential method described 
above. The hyperfine Hamiltonian is not derived via the Breit equation and 
hence the FSC are not obtained as perturbations to the Breit equation.
Zemach starts directly deriving the hyperfine energy shift $\Delta E_Z$ by
introducing a magnetic field ${\bf H}({\bf r})$ such that,   
\begin{equation}\label{zemacheq1}
\Delta E_Z = \mu_1 \int\, \Psi_Z^*({\bf r}) \langle {\bm \sigma}_2 \cdot 
{\bf H}({\bf r})\rangle 
\Psi_Z({\bf r}) 
\end{equation}
where part of the FSC has already been incorporated in the wave function 
$\Psi_Z$ which is a solution of the Sch\"odinger equation 
for an electron moving in the field of the distribution $e f_e({\bf r})$. 
$\mu_1$ denotes the electron magnetic moment and $f_e({\bf r})$ is the 
Fourier transform of one of the proton's form factors $F_1(q^2)$.
The magnetic field ${\bf H}({\bf r})$ at a given point {\bf r} 
which is usually the field due to a point magnetic dipole is then modified 
to include the Fourier transform of the magnetic Sachs form factor $G_M(q^2)$ 
mentioned earlier in the article. As a result, Zemach finally obtains, 
\begin{equation}\label{zemacheq2}
\Delta E_Z = - {2 \over 3} \mu_1 \,\mu_2 \, \langle {\bm \sigma}_1  
\cdot {\bm \sigma}_2 \rangle \, \int\, |\Psi_Z({\bf r})|^2\, f_m({\bf r})\, 
d{\bf r}
\end{equation}
where $\mu_{1,2} = e_{1,2}/2m_{1,2}$ and $f_m({\bf r})$ is the Fourier 
transform of $G_M(q^2)$. 

A comparison of 
Eqs (\ref{zemacheq1}) and (\ref{zemacheq2}) with Eq. (\ref{encorr}) 
makes it obvious that the Zemach and Breit approaches are completely 
different. Whereas the Breit approach involves the expectation value 
(using unperturbed wave functions) of 
a hyperfine potential with form factors, the Zemach approach has the 
proton structure information embedded in the wave function $\Psi_Z({\bf r})$ 
(in terms of the $F_1(q^2)$ form factor) as well as the magnetic field 
(in the form of $G_M(q^2)$). 

In the same work, following a perturbative formalism as given by 
Karplus and Klein, Zemach obtains another expression for $\Delta E_Z $ 
(Eq. (5.8) in \cite{zemach}) 
which apart from a small difference in the use of the reduced mass 
is nothing but the momentum space representation of (\ref{zemacheq2}). 
Performing an explicit comparison of this expression 
with (\ref{encorr}) 
by rewriting (\ref{encorr}) in momentum space using (\ref{potqhfs}) it is 
easy to see that there arise some additional terms in the Breit method as 
compared to the Zemach approach.

Zemach further introduced some approximations and a new distribution function 
$f_{em}({\bf r}) = \int f_e({\bf r} - {\bf s}) f_m({\bf s}) d{\bf s}$ to 
obtain the well known Zemach formula, 
$\Delta E_Z = \Delta E_0 (1 - 2m_1 \alpha \langle r \rangle _{em})$ where 
$\langle r \rangle _{em} = \int r f_{em}({\bf r}) d{\bf r}$.  

\subsubsection{Bodwin and Yennie's correction}
In \cite{yennie}, in addition to computing the recoil corrections to the 
hydrogen hyperfine splitting, the authors provided a formalism to obtain 
the proton structure corrections to the hyperfine splitting in hydrogen. 
The theoretical procedure involves the evaluation of perturbation kernels 
corresponding to structure dependent photon-proton vertices. A given 
perturbation kernel does not yield a result of unique order in $\alpha$ but 
the kernels are ranked in importance according to the largest contribution 
that they produce. It was found that it was sufficient to include the 
contribution of the one loop kernel to the hfs. The six term lengthy 
expression for $\Delta E_{hfs}$ was analysed and the first term 
after some approximations was 
found to reproduce the Zemach formula. Thus the Zemach correction within 
the formalism of Bodwin and Yennie was given as,
\begin{equation}\label{yennie}
\Delta E = E_F {2 \alpha m_r \over \pi^2} \, \int d^3p \, 
{1 \over ({\bf p}^2 + \gamma^2)^2} \, \biggl [ {G_E(-{\bf p}^2) 
G_M(-{\bf p}^2)  \over 1 + \kappa} - 1 \biggr ]\, ,
\end{equation}
where $E_F$ is the Fermi energy and 
we refer the reader to \cite{yennie} for details of the notation. 
The authors found that the net contribution of the remaining five terms 
in their expression was small as compared to the Zemach correction. 
A comparison of the results of the present work and corrections 
obtained using other methods is presented at the end of this section. 

A related approach 
was presented in \cite{pach} where the Zemach correction was reproduced 
in the limit of large proton mass. The 
order $\alpha^5$ corrections to the hyperfine energy levels 
with proton structure included were evaluated from two photon 
exchange amplitudes. In Zemach's 
original work both one and two photon exchanges between the electron and 
proton were considered. Zemach noted that for the two photon exchange case the 
interaction is of second order in $\alpha$ (i.e. one order higher than the 
one photon exchange interaction) and that the contribution of this term is 
very small. 
The Breit potential used in
time-independent perturbation theory gives the 
$\alpha^5$ and higher order corrections to the energy 
from the one photon exchange diagram (this potential is 
first order in $\alpha$ as in Zemach's work) . 
Clearly approaches such as in \cite{pach} will differ from the Breit method 
and lead to different results for the FSC.

\subsubsection{Order $\alpha^5$ corrections with one and two photon exchange} 
Note that the corrections obtained in the present work are the
finite size corrections to the hyperfine part of the Breit 
Hamiltonian. Once the $q^2$ dependent form factors are replaced by those 
at $q^2 =0$, one recovers the standard 
hyperfine Hamiltonian for point particles. 
This is different from some of the methods mentioned above which evaluate the 
finite size and recoil corrections within the same formalism. 
They recover the recoil correction to the hyperfine structure 
on substituting form factors at $q^2 = 0$ in the full expression (which 
involves the Zemach plus recoil corrections) \cite{pach}.
 
We emphasize here that the approach in the present work is 
conceptually quite different from the ones above which are 
based on the Bethe Salpeter equation. There one starts by writing the 
hyperfine energy contribution (at order $\alpha^5$) 
induced by a skeleton diagram with two photon exchange. This consists of 
a divergent integral, namely, 
$$ {8 Z \alpha \over \pi n^3} E_F \, \int \, {dk \over k^2} $$
where $E_F$ is the Fermi energy and is of order $\alpha^4$. Insertion of the 
electromagnetic form factors leads to the correction (Zemach term of 
Bodwin and Yennie), 
\begin{equation}
\Delta E = {8 Z \alpha m \over \pi n^3} E_F \, \int \, {dk \over k^2} 
\, \biggl [ (G_E(-k^2) - 1) + \biggl ( {G_M(-k^2) \over 1 + \kappa} - 1 
\biggr )\, \biggr ] \, .
\end{equation}
Note that the introduction of finite size effects through the electromagnetic 
form factors $G_E$ and $G_M$ has not changed the order of $\alpha$ in the 
energy (i.e. to say that the order $\alpha^5$ is due to two photon 
exchange and not due to the introduction of finite size effects). 
In the Breit potential approach of the present work, the hyperfine energy 
calculated using the potential for point-like protons is of order $\alpha^4$ 
(this is the Fermi energy). The introduction of the potential with form 
factors (which is of the same order in $\alpha$ as the point potential)
to evaluate the energy using first order perturbation theory 
leads to order $\alpha^5$ and higher order corrections terms from the 
same one photon exchange diagram. The higher orders in $\alpha$ arise due 
to the perturbation theory approach and not due to more photons being 
exchanged. This is evident from Eqs (\ref{energyhfs12}) where the general 
expression for the hyperfine energy level for any $n$ and $l =0$ is 
written in a more lucid way by replacing a series expansion of the 
confluent hypergeometric functions. 
One can see (using $a = 1/(\alpha m_r)$)  
that order $\alpha^5$ and higher corrections appear in the 
same expression obtained from a one photon exchange diagram. 
Setting $m \to \infty$ 
in Eqs (\ref{energyhfs12}) in order to 
remove the effect of proton structure returns the order $\alpha^4$ 
Fermi energy.

\subsubsection{Friar's finite size correction} 
In passing, we finally note that there exist some other approaches such as the 
one proposed by Friar \cite{friar}. The finite size correction within this 
approach seems to be independent of the parametrization of the form factors. 
However, the hydrogen atom wave function is not taken to be ${\bf r}$ 
dependent but rather at ${\bf r} =0$. Besides, it also appears that though 
the charge distribution of the proton is taken into account, the magnetization 
distribution is omitted. 

We end this section by mentioning that 
the correction of -58.2 kHz in \cite{friar} evaluated to the 1S
hyperfine level is very close to -60.2 kHz obtained in \cite{eides}. It 
differs from the correction of the present work but is of 
the same order of magnitude. 
Within the Breit potential approach we 
obtain the FSC of 
-37.696 kHz to the 1S level splitting in hydrogen. 
Changing the FSC of -60.2 kHz to our value -37.696 kHz
would change the total theoretical hfs value as calculated in \cite{eides} 
(Table 19) from 1420399.3 kHz to 1420421.8 kHz. This is 
important considering the exact experimental value 1420405.751768(1) kHz 
and the fact that radiative corrections range in the order
$10^{-2}$-$10^{-3}$ kHz \cite{eides} (Tables 14-15).
The FSC to  
$D_{21} = 8\, E^{2S}_{hfs} \, -\,  E^{1S}_{hfs}$
is also sensitive to the method used. In \cite{karshenPLB}, 
the correction, $D_{21} (nucl) = -0.002$ kHz as 
compared to $-0.0000833$ kHz of the present work. Such a small correction
makes $D_{21}$ almost free of finite size effects. The 
experimental value for $D_{21}$ is 48.923(54) kHz \cite{karshnew}. Fourth
order QED corrections are 0.018 kHz \cite{karshnew} and 
comparable to -0.002 kHz, however, bigger than our estimate of FSC. 
From Eqs (\ref{energyhfs12}) it is obvious that the FSC to $D_{21}$ begin 
at order $\alpha^6$ and hence are expected to be very small. 

\subsection{Hyperfine splitting in muonic hydrogen} 
In Table II we present the 
corrections to the energies 
calculated using Eq. (\ref{energyhfs12}) but for the case of 
muonic hydrogen ($\mu^- p$) 
2S splitting and Eq. (\ref{energyhfs2p}) for the 2P$j$ ($j = 1/2, 3/2$) 
splitting in
$\mu^- p$. The individual level energies are also listed. The 
splitting 2S is the difference 2S$_{1/2}^{f=1}$ - 2S$_{1/2}^{f=0}$ and 
the splittings 2P$_{1/2}$ and 2P$_{3/2}$ 
are 2P$_{1/2}^{f=1}$ - 2P$_{1/2}^{f=0}$ and 
2P$_{3/2}^{f=2}$ - 2P$_{3/2}^{f=1}$ respectively. In fact, 
the energy of the 2S splitting is 4 E$_{hfs}$(2S$_{1/2}^{f=1}$).  
\begin{table}
\caption{ Energies in meV of hyperfine levels and 
splittings in muonic hydrogen atom.} 
\label{tab:2}
\begin{tabular}{|c|c|c|c|}\hline
Level &$E^0$  &$\Delta E^{FSC}$& $\Delta E^{corr}$ \\
     &       & due to proton structure & 
due to $\kappa_{\mu}$ and proton structure\\
\hline
2S$_{1/2}^{f=1}$ & 5.70135  &-0.0280860  & -0.02144 \\ \hline
2P$_{1/2}^{f=1}$ & 1.90045  & -0.0000064  &  0.00110 \\ \hline
2P$_{1/2}^{f=0}$ &-5.70135  & 0.0000345  & -0.00329 \\ \hline
2P$_{3/2}^{f=2}$ & 1.14027  & -0.0000077  & -0.00034  \\ \hline
2P$_{3/2}^{f=1}$ &-1.90045  & 0.0000103  &  0.00056 \\ \hline
Splitting &   & &  \\ \hline
2S          &22.80541   & -0.112342  & -0.085753   \\ \hline
2P$_{1/2}$  & 7.60180   & -0.000041  &  0.004390 \\ \hline
2P$_{3/2}$ &  3.04072 &   -0.000018  & -0.000904  \\ \hline
\end{tabular} 
\end{table}
In the evaluation of the proton radius in 
\cite{naturerp}, the values of the hyperfine splittings were taken from
\cite{martynenkos}, where the FSC for the 2S were evaluated using the 
Zemach method and those for the 2P case were not taken into account. 
Their FSC (taken from Table II of the first reference in \cite{martynenkos}) 
of order $\alpha^5$ and $\alpha^6$ sum to -$0.1535$ meV in contrast to the 
-$0.11234$ meV of the present work. This result would change the input of 
$\Delta E_{HFS}^{2S}$ = 22.8148 meV (which includes the FSC 
corrections using Zemach's method) used in \cite{naturerp} to 
$\Delta E_{HFS}^{2S}$ = 22.8560 meV. Correcting the 2P hyperfine splitting 
in \cite{naturerp}, $\Delta E_{HFS}^{2P_{3/2}}$ = 3.392588 meV would change 
to $\Delta E_{HFS}^{2P_{3/2}}$ = 3.392570 meV. These corrections are quite 
relevant considering the precision taken into account in \cite{naturerp} 
while deducing the charge radius of the proton. 

\section{Summary} 
To summarize, the Breit potential for hydrogen atom 
with the inclusion of electromagnetic form factors of the proton 
is presented. This includes the proton structure corrections to 
the standard Coulomb potential plus terms 
such as the Darwin term (which takes into account the relativistic effects), 
the spin-spin and spin-orbit interaction terms (corresponding to 
fine and hyperfine structure) and retarded potential terms. We focused in 
particular on certain terms involving the spin-spin and spin-orbit interaction 
with the aim of studying the proton 
form factor effects (or proton finite size effects) in the hyperfine splitting 
in electronic and muonic hydrogen. 
The finite size corrections (FSC) 
to the potentials originate from the way the photon 
interacts with extended objects. It is therefore possible to include such 
corrections already in the standard Breit equation where the potential is 
of order $\alpha$.
It is important to note that the 
standard hyperfine Hamiltonian for hydrogen atom follows naturally from 
the expressions for the Breit potential with form factors, once the 
form factors are replaced by their values at $q^2 = 0$. This Hamiltonian 
contains $\kappa_p (= F_2^p(0))$, 
the anomalous magnetic moment of the proton which is 
large due to the fact that the proton is an extended object. 
Hence part of the finite size correction is already inherent in the 
hyperfine Hamiltonian. By including $q^2$ dependent form factors, we just 
make the Hamiltonian more general and complete. 
We presented here
the FSC for three cases: 
1S and 2S splittings in hydrogen atom, Sternheim's observable $D_{21}$ and
2S and 2P splittings in muonic hydrogen.  
The theoretical aspects as well as numerical values obtained in the present 
work were compared with other existing methods for FSC in literature.
The present work aims also at showing that there exist 
different sophisticated and consistent approaches to evaluate the finite 
size corrections.  

Our FSC of -$0.11234$ meV to the 2S hyperfine splitting in muonic hydrogen 
is close to the order $\alpha^5$ corrections of -$0.183$ meV \cite{pach} and 
-$0.1518$ meV \cite{martynenkos} using other approaches. As one can see, 
though the three numbers are of the same order of magnitude, they are not 
equal to each other. Thus, one can say that there is a small uncertainty 
in the calculation of the finite size effects, introduced due to the 
differences in the approaches used for FSC.

The present work gives the analytical expressions for the hyperfine 
potential and energy 
levels including the finite size corrections due to the structure of the 
proton. This has been achieved using the dipole form of the proton form 
factors. Note that the inclusion of proton structure effects gives rise
to two very different 
length scales in the same calculations: the atomic one involving 
the Bohr radius $a$ (which is of the order of 10$^4$ fm)
and the nuclear one with $r_p \propto 1/m$ (of the order of 1 fm). 
This would make a numerical evaluation of the 
integrals involved in the calculations tedious and hence an analytical 
approach as in the present work is preferable.
\renewcommand{\theequation}{A-\arabic{equation}}
\setcounter{equation}{0}
\section*{APPENDIX: Hyperfine energy levels with proton structure 
corrections for any $n$ and $l \ne 0$}

The hyperfine energy for any $n$ and $l \ne 0$ is evaluated by taking the 
expectation value of the potential in Eq. (\ref{potrhfs}). We start 
by rewriting the potential as follows for convenience: 
\begin{eqnarray}
&&\hat{V}_{hfs}=\frac{\alpha}{m_em_pc^2}(1+\kappa_e)(1+\kappa_p)\Bigg
[3(\textbf{S}_e.\hat{\textbf{r}})(\textbf{S}_p.\hat{\textbf{r}})
\bigg(\frac{1}{r^3}+\frac{{h}_{1}}{r^3}\bigg)-
(\textbf{S}_e.\textbf{S}_p)\bigg(\frac{1}{r^3}+\frac{{h}_2}{r^3}\bigg)
\Bigg]\nonumber\\
&& +\frac{e^2}{m_em_pc^2}(\textbf{L}.\textbf{S}_p)\Bigg[(1+\kappa_p)\left
(\frac{1}{r^3}+\frac{{h}_3}{r^3}\right)+\frac{m_e}{2m_p}\frac{{h}_4}{r^3}
\Bigg],\label{4_31}
\end{eqnarray}
where,  
\begin{eqnarray}
 {h}_{1}&=
&-e^{-mr}(1+mr)-\frac{m^2r^2}{2}e^{-mr}-\frac{m^3r^3}{6}e^{-mr}\nonumber\\
 {h}_{2}&=&-e^{-mr}(1+mr)-\frac{m^2r^2}{2}e^{-mr}-\frac{m^3r^3}{2}e^{-mr}
\nonumber\\
 {h}_{3}&=&-e^{-mr}(1+mr)-\frac{m^2r^2}{2}e^{-mr}\nonumber\\
 {h}_{4}&=&(1+2\kappa_p)(1+{h}_{3})+
\frac{\kappa_p}{(1-k^2)^2}e^{-mr}(1+mr)+\frac{\kappa_p}{1-k^2}
\frac{m^2r^2}{2}e^{-mr}\nonumber\\&&-
\frac{\kappa_p}{(1-k^2)^2}e^{-mkr}(1+mkr),\nonumber
\end{eqnarray}
and $k=2m_p/m$. 
Eq. (\ref{4_31}) was obtained using the identity, 
\begin{equation}
 (\hat{\bm{\sigma}}_e.\vec{\nabla})(\hat{\bm{\sigma}}_p.\vec{\nabla})G(r)=(\hat{\bm{\sigma}}_e.\hat{\bm{\sigma}}_p)\left(\frac{1}{r}\frac{\partial G(r)}{\partial r}\right)+\frac{(\hat{\bm{\sigma}}_e.\textbf{r})(\hat{\bm{\sigma}}_p.\textbf{r})}{r^2}\left(\frac{\partial^2 G(r)}{\partial r^2}-\frac{1}{r}\frac{\partial 
G(r)}{\partial r}\right).\label{A41}
\end{equation}
Using first order time independent perturbation theory, 
$$E_{hfs}^{nl} = \int\, \Psi_{nlm}^*({\bf r})\, \hat{V}_{hfs}\,\Psi_{nlm}({\bf r}) 
d{\bf r} , $$
where, 
\begin{equation}
\Psi_{nlm}(r,\theta,\phi)= \Bigg[\left(\frac{2}{na}\right)^3\frac{(n-l-1)!}{2n(n+l)!}\Bigg]^{1/2}e^{-r/na}\left(\frac{2r}{na}\right)^lL_{n-l-1}^{2l+1}(2r/na)Y_l^{m}(\theta,\phi) \label{B6}
\end{equation}
is the wave function of the hydrogen atom. 
The identity, 
$$
\left[L_{n-l-1}^{2l+1}\biggl({2r\over na}\biggr)
\right]^2=\frac{\Gamma(n+l+1)}{2^{2(n-l-1)}(n-l-1)!}
\sum_{j=0}^{n-l-1}{2(n-l-j-1)\choose n-l-j-1}     
\frac{(2j)!}{j!\Gamma(2l+j+2)}L_{2j}^{2(2l+1)}\biggl({4r\over na}\biggr),
$$
allows one to write the expectation value of a function sandwiched between 
the hydrogen wave functions as,
\begin{eqnarray}
 \langle A\rangle&=&\left(\frac{2}{na}\right)^{2l+3}\frac{1}{2n2^{2(n-l-1)}}\sum_{j=0}^{n-l-1}{2(n-l-j-1)\choose n-l-j-1}\frac{(2j)!}{j!\Gamma(2l+j+2)}  \nonumber\\&&
 \times\int_0^\infty dr A(r) e^{-2r/na}r^{2l+2}L_{2j}^{2(2l+1)}(4r/na).\label{B9}
\end{eqnarray}
This leads to, 
\begin{eqnarray}
&&E_{hfs}^{nl}\left(j,f,\left<\textbf{S}_e.\textbf{S}_p\right>\right)=
\frac{e^2}{m_em_p}\Bigg[\Bigg(-(1+\kappa_e)(1+\kappa_p)\left(j(j+1)-l(l+1)-
\frac{3}{4}\right)\nonumber\\&&
\times\left<\frac{1}{r^3}+\frac{{h}_{1}}{r^3}\right>+
\left(j(j+1)+l(l+1)-\frac{3}{4}\right)\left<(1+\kappa_p)\left(\frac{1}{r^3}+
\frac{{h}_3}{r^3}\right)+\frac{m_e}{2m_p}\frac{{h}_4}{r^3}\right>
\Bigg)\nonumber\\&&
\times\left(\frac{f(f+1)-j(j+1)-\frac{3}{4}}{4j(j+1)}\right)+\frac{m^3}{3}
\left<\textbf{S}_e.\textbf{S}_p\right>\left<e^{-mr}\right>\Bigg],\label{4_35}
\end{eqnarray}
where, the expectation values are: 
\begin{eqnarray}
 &&\left<\frac{1}{r^3}+\frac{{h}_{1}}{r^3}\right>=\left(\frac{2}{na}\right)^
{2l+3}\frac{1}{2n2^{2(n-l-1)}\Gamma(4l+3)}\sum_{j=0}^{n-l-1}{2(n-l-j-1) \choose n-l-j-1}\frac{\Gamma(4l+2j+3)}{j!\Gamma(2l+j+2)}\nonumber\\&&
 \Bigg[\Gamma(2l)\left(\frac{na}{2}\right)^{2l} \,_2F_1\left(-2j,2l;4l+3;2\right)-
\Gamma(2l)\left(\frac{na}{2+mna}\right)^{2l} \,_2F_1\left(-2j,2l;4l+3;
\frac{4}{2+mna}\right)\nonumber\\&&
 -m\Gamma(2l+1)\left(\frac{na}{2+mna}\right)^{2l+1} \,_2F_1\left(-2j,2l+1;4l+3;\frac{4}{2+mna}\right)	-\frac{m^2}{2}\Gamma(2l+2)	\nonumber\\&&
 \left(\frac{na}{2+mna}\right)^{2l+2}\, _2F_1\left(-2j,2l+2;4l+3;\frac{4}{2+mna}\right)-\frac{m^3}{6}\Gamma(2l+3)\left(\frac{na}{2+mna}\right)^{2l+3}\nonumber\\&&
 \,_2F_1\left(-2j,2l+3;4l+3;\frac{4}{2+mna}\right)\Bigg]\label{4_36},
\end{eqnarray}
\begin{eqnarray}
&&\left<(1+\kappa_p)\left(\frac{1}{r^3}+\frac{{h}_3}{r^3}\right)+
\frac{m_e}{2m_p}\frac{{h}_4}{r^3}\right>=\left(\frac{2}{na}\right)^{2l+3}
\frac{1}{2n2^{2(n-l-1)}\Gamma(4l+3)}\nonumber\\&&
 \sum_{j=0}^{n-l-1}{2(n-l-j-1) \choose n-l-j-1}
\frac{\Gamma(4l+2j+3)}{j!\Gamma(2l+j+2)}\Bigg[\left(1+\kappa_p+\frac{m_e}{2m_p}(1+2\kappa_p)\right)\Gamma(2l)\left(\frac{na}{2}\right)^{2l}\nonumber\\&&
\, _2F_1\left(-2j,2l;4l+3;2\right)-\left(1+\kappa_p+\frac{m_e}{2m_p}\left(1+2\kappa_p-\frac{\kappa_p}{(1-k^2)^2}\right)\right)\Gamma(2l)\left(\frac{na}{2+mna}\right)^{2l}\nonumber\\&&
\, _2F_1\left(-2j,2l;4l+3;\frac{4}{2+mna}\right)-\left(1+\kappa_p+\frac{m_e}{2m_p}\left(1+2\kappa_p-\frac{\kappa_p}{(1-k^2)^2}\right)\right)m\Gamma(2l+1)\nonumber\\&&
 \left(\frac{na}{2+mna}\right)^{2l+1} \,_2F_1\left(-2j,2l+1;4l+3;\frac{4}{2+mna}\right)-\left(1+\kappa_p+\frac{m_e}{2m_p}\left(1+2\kappa_p-\frac{\kappa_p}{1-k^2}\right)\right)\nonumber\\&&
 \frac{m^2}{2}\Gamma(2l+2)\left(\frac{na}{2+mna}\right)^{2l+2} \,_2F_1\left(-2j,2l+2;4l+3;\frac{4}{2+mna}\right)-\frac{m_e}{2m_p}\frac{\kappa_p}{(1-k^2)^2}\Gamma(2l)    \nonumber\\&&
 \left(\frac{na}{2+mkna}\right)^{2l} \, _2F_1\left(-2j,2l;4l+3;\frac{4}{2+mkna}\right)-\frac{m_e}{2m_p}\frac{\kappa_pmk}{(1-k^2)^2}\Gamma(2l+1)  \left(\frac{na}{2+mkna}\right)^{2l+1}   \nonumber\\&&
\,_2F_1\left(-2j,2l+1;4l+3;\frac{4}{2+mkna}\right)\Bigg],\label{4_37}
\end{eqnarray}
and 
 \begin{eqnarray}
  \left\langle e^{-mr} \right\rangle&=&\left(\frac{2}{na}\right)^{2l+3}\frac{1}{2n2^{2(n-l-1)}}\sum_{j=0}^{n-l-1}{2(n-l-j-1)\choose n-l-j-1}\frac{\Gamma(4l+2j+3)}{j!\Gamma(2l+j+2)}  \nonumber\\&&
  \frac{\Gamma(2l+3)}{\Gamma(4l+3)}\left(\frac{na}{2+mna}\right)^{2l+3}
\,_2F_1\left(-2j,2l+3;4l+3;\frac{4}{2+mna}\right) . \label{B12}
 \end{eqnarray}

\end{document}